\documentclass[
,final
]{aipproc}
\layoutstyle{6x9}

\begin{document}

\title{
\begin{flushright}\vspace{ -.1in}
{\normalsize MICE-CONF-GEN-121\\
\vspace{ -.05in}
}
\end{flushright}
MICE: The International Muon Ionization Cooling Experiment}

\author{Daniel M.\ Kaplan}{address={Illinois Institute of Technology, Chicago, Illinois 60616, USA\\[.15in]
{\rm (for the MICE Collaboration)}}}
\begin{abstract}
Ionization cooling of a muon beam is a key technique for a Neutrino Factory or Muon Collider.
An international collaboration is mounting an experiment to demonstrate muon ionization cooling at the Rutherford Appleton Laboratory. We aim to complete the experiment by 2010.
\end{abstract}

\maketitle
\section{introduction}
The experimental establishment of neutrino oscillations~\cite{nu-osc} has stimulated widespread interest in a muon storage ring-based Neutrino Factory~\cite{Geer}, possibly the ultimate tool for studying the neutrino mixing matrix~\cite{ultimate}. Two feasibility studies~\cite{feas1,feas2} have shown that a high-performance Neutrino Factory can be built using available technology. However, some of the beam-manipulation techniques envisaged have yet to be applied in practice. 
Of these, ionization cooling of the muon beam~\cite{Skrinsky-etal,Neuffer-Fernow} is perhaps the most novel. In the longer term it holds the promise of $s$-channel Higgs Factories and multi-TeV muon-antimuon colliders, with potential for unique studies of matter and energy at the most fundamental level~\cite{mucollider}, complementing those at the Large Hadron Collider and the proposed International Linear Collider. 

Ionization cooling contributes significantly to both the performance (up to a factor of 10 in 
intensity~\cite{Hanke}) and cost (as much as 20\%~\cite{feas2}) of a Neutrino Factory. This 
motivates 
the Muon Ionization Cooling Experiment (MICE). MICE is intended not only to demonstrate the {\em principle} of ionization cooling, but (and perhaps more importantly) to show how to build and operate a device with the performance required for a Neutrino Factory. 
The experience gained from MICE will provide input to the final design of the Neutrino Factory cooling channel and firm up its cost estimate. An important part of the MICE program is to study the cooling process by varying the relevant 
parameters, so that an extrapolation can be made to a different cooling-channel design, e.g., a ring~\cite{Palmer-ring} or helical cooling channel~\cite{Derbenev}, should one of these be shown to be advantageous. 

During 2001 and 2002, the international MICE Collaboration~\cite{MICE} was formed and developed a proposal~\cite{MICE-proposal} to carry out this program using a muon beam produced with the ISIS accelerator at  Rutherford Appleton Laboratory (RAL). The proposal was approved in 2003. 
The MICE collaboration includes accelerator  and experimental particle physicists from Europe, Japan, and the US. As of this writing, funding for the first phase of MICE has been provided in Italy, Japan, the Netherlands, Switzerland, the UK, and the US. We aim for a definitive demonstration of ionization cooling by 2010.

\section{Design of the experiment}
The MICE design is presented in detail in the proposal~\cite{MICE-proposal} and Technical Reference Document (TRD)~\cite{TRD}, and is briefly summarized here.  
The goals of MICE are 
\begin{itemize}
\item to engineer and build a section of cooling channel (of a design that can give the desired performance for a Neutrino Factory) that is long enough to provide a measurable ($\approx$10\%) cooling effect, but short enough to be moderate in cost;  and
\item to 
measure the resulting cooling effect with an absolute accuracy of 0.1\%  
over a muon-beam momentum range of 140--240\,MeV/$c$. 
\end{itemize}

\begin{figure}[hbt]
\hspace{.1in}\centerline{\scalebox{0.75}{\includegraphics*[bb=0 433 650 750,clip]{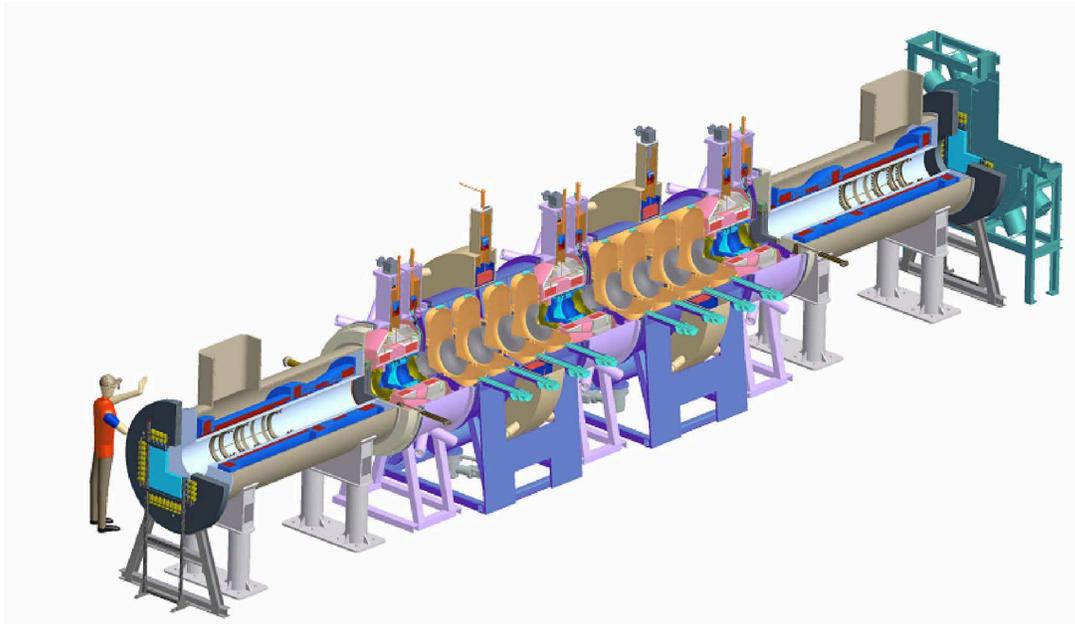}}}
\caption{Three-dimensional cutaway rendering of the MICE apparatus. The muon beam enters from the lower left and is measured by time-of-flight (TOF) and Cherenkov detectors and a first solenoidal tracking spectrometer. It then enters the cooling section, where it is alternately slowed down in absorbers and reaccelerated by rf cavities, while being focused by a lattice of superconducting solenoids. Finally it is remeasured by a second solenoidal tracking spectrometer and its muon identity confirmed by Cherenkov and TOF detectors and a calorimeter.
}
\label{fig:MICE}
\end{figure}

The layout of MICE is shown in Fig.~\ref{fig:MICE}. The tracks of single muons through the apparatus will be measured using standard particle-physics techniques, since bunched-beam diagnostics lack the needed precision. 
The 5.5\,m-long cooling section, consisting of three absorbers and eight rf cavities encircled by lattice solenoids, is therefore surrounded at each end by tracking detectors, to measure beam emittance at the entrance and exit, and particle-ID detectors to reject particles other than muons. This requires the placement of tracking detectors close to the rf cavities and is therefore sensitive to backgrounds caused by dark-current electrons and their associated x-rays. Improving our understanding of such backgrounds is essential to the successful planning and execution of MICE and is a much-anticipated result from upcoming tests by the MuCool Collaboration~\cite{MuCool-rf}.

The cooling section is one lattice cell of the ``SFOFO'' cooling channel developed in Feasibility Study II~\cite{feas2} (with minor modifications to reduce cost and comply with RAL safety requirements), with a full absorber at each end to 
protect the detectors from rf-cavity emissions. This arrangement provides considerable flexibility, as the solenoid polarities and currents can be varied to test a variety of lattices. Provision will be made 
for a variety of solid absorbers as well as liquid hydrogen and helium. (In principle, some other cooling cell could also be tested, perhaps in a subsequent MICE phase.)

\begin{figure}
\hspace{.2in}\scalebox{.4}{\includegraphics*[bb=20 240 550 475,clip]{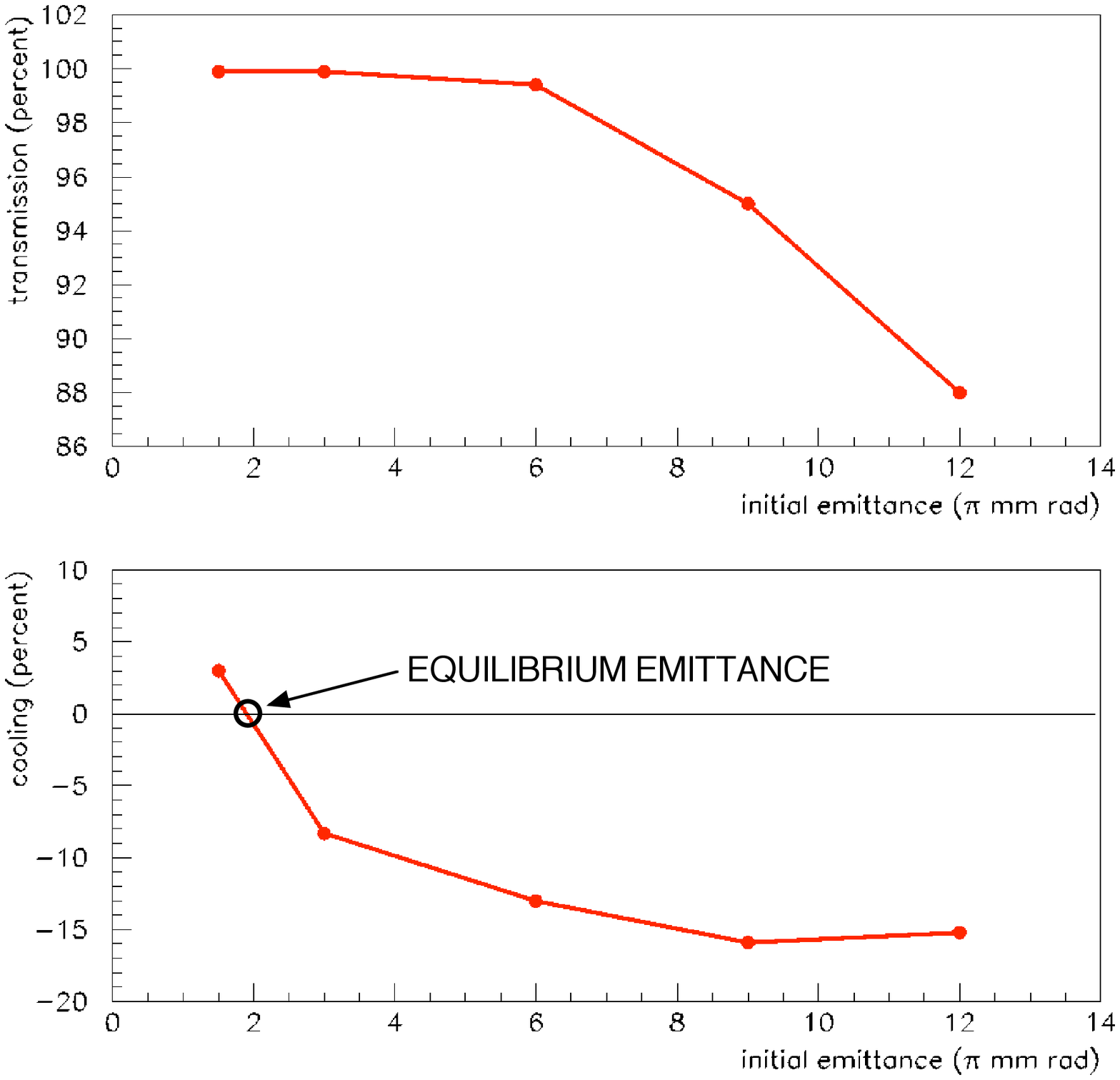}\includegraphics*[bb=20 493 595 728,clip]{MICE-perf2}}
\caption{(Left) percent change of normalized transverse emittance and (right) beam transmission through cooling section, both vs.\ input emittance in $\pi$\,mm$\cdot$rad.}\label{fig:perf}
\end{figure}

Figure~\ref{fig:perf} shows the simulated effect of the cooling section on the normalized transverse beam emittance, as well as the beam transmission,  vs.\ that emittance, for 200\,MeV/$c$ average beam momentum and nominal optics settings of the SFOFO lattice cell (3.2\,T maximum on-axis field). For input emittance below the equilibrium value of $2\pi$\,mm$\cdot$rad the beam is heated; above $6\pi$\,mm$\cdot$rad scraping begins to deplete the beam. The detailed comparison of such measurements against Monte Carlo predictions, for a variety of beam momenta, emittances, and apparatus configurations, will serve to validate our Monte Carlo and design approach and allow extrapolation to the longer ($\sim$100\,m) cooling channels typically used in Neutrino Factory designs.

\begin{figure}
\centerline{\scalebox{1.1}{\includegraphics*[bb=200 330 400 473,clip]{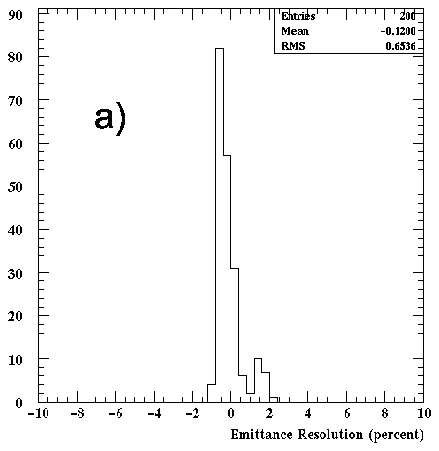}}\hspace{-.8in}\scalebox{1.15}{\includegraphics*[bb=203 333 400 460,clip]{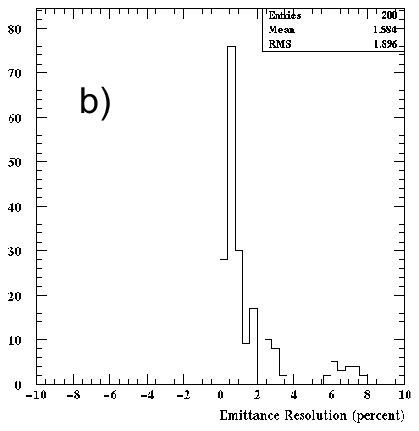}}}
\caption{Resolution in uncorrected emittance (for 2.5$\pi$\,mm$\cdot$rad input emittance) in a)~upstream and b) downstream spectrometer.}\label{fig:Emitt-res}
\end{figure}

Achieving 0.1\% emittance resolution will require careful calibration and simulation. 
Scattering of the beam in the detectors causes a correctable bias, as illustrated (for input-beam emittance $\varepsilon_t=2.5\pi$\,mm$\cdot$rad) in Fig.~\ref{fig:Emitt-res}~\cite{Khan}: 
before correction, the transverse emittance measured in each spectrometer is $\sim$1\% larger than the ``true'' emittance. The goal of 0.1\% emittance measurement will thus require that this bias be calibrated and corrected to $\sim$10\% of itself (to be verified by calibration runs with no cooling section).

\subsection{Tracking detectors}
To minimize beam scattering and sensitivity to x-rays, the tracking detectors will be thin (350\,$\mu$m-diameter) scintillating fibers, ganged by sevens to reduce the needed electronics channel count. Each group of seven adjacent fibers is mated to a 1\,mm clear light-guide fiber that conveys the scintillation light to a VLPC photosensor. The $>$85\% quantum efficiency of the VLPCs~\cite{D0-SciFi} results in an average of 11 photoelectrons per minimum-ionizing particle, as verified in cosmic-ray tests~\cite{Khan}. As in the D0 experiment~\cite{D0-SciFi}, the use of two staggered layers per view ensures high efficiency. Each spectrometer will be made up of five detector stations, each with three views arranged in 120$^\circ$ stereo, deployed within a 1.1\,m-long  4\,T superconducting solenoid. A prototype 4-station detector is now undergoing beam tests in a 1\,T solenoid at KEK.

\subsection{Particle identification}
The muon beam may contain residual pions which are transported through the large momentum acceptance of the beamline, as well as electrons from the in-flight decay of muons. A three-plane time-of-flight system provides the precise time information needed for particle identification, emittance measurement, and off-line bunch construction and timing with respect to the rf phase. Additional particle identification is provided before and after the cooling channel by Cherenkov detectors and a calorimeter.

\section{Staging and current status}
\begin{figure}
\scalebox{.5}{\includegraphics*[bb=0 350 600 760,clip]{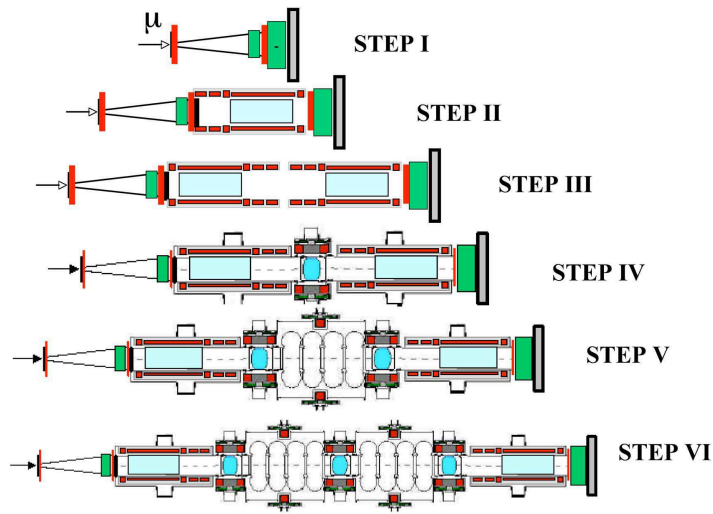}}
\caption{Six possible steps in the development of MICE.}\label{fig:steps}
\end{figure}

The need to carefully cross-calibrate the spectrometers as well as the cost of the cooling section suggest staging the installation and operation of MICE as indicated in Fig.~\ref{fig:steps}. The currently funded first phase of MICE includes the detectors but not the cooling section; the second phase will be assembled in steps as  funds allow. 
At present  the first rf cavity is under test in the MuCool Test Area at Fermilab, the Absorber/Focus-Coil  and RF-cavity/Coupling-Coil module designs are well advanced, and work is
in progress on the spectrometers, beamline, and infrastructure. First beam is planned for April 2007.

\section{Acknowledgements}

The author gratefully acknowledges support of the US MICE institutions by the Department of Energy and the National Science Foundation.

\end{document}